\documentclass[%
 reprint,
superscriptaddressx
 amsmath,amssymb,
 aps,prl]{revtex4-2}

\usepackage{blindtext}
\usepackage{natbib}
\usepackage{multirow}
\usepackage{graphicx}
\usepackage{dcolumn}
\usepackage{bm}
\usepackage{hyperref}
\usepackage{cleveref}
\usepackage{siunitx}
\usepackage{xspace}

\Crefname{figure}{\text{Fig.}}{\text{Figs.}}

\newcommand{\SiOO}{SiO$_{2}$\xspace}
\newcommand{\SiOSi}{SiO$_{2}$/Si\xspace}

\newcommand{\LEmuSR}{LE-$\mu$SR\xspace}
\newcommand{\muSR}{$\mu$SR\xspace}
\newcommand{\MuBCz}{Mu$^\textrm{0} _\textrm{BC}$\xspace}
\newcommand{\MuBC}{Mu$_\textrm{BC}$\xspace}
\newcommand{\MuBCp}{Mu$^{+} _\textrm{BC}$\xspace}

\newcommand{\MuTz}{Mu$^\textrm{0} _\textrm{T}$\xspace}
\newcommand{\MuTm}{Mu$^{-}_\textrm{T}$\xspace}
\newcommand{\FD}{F$_\textrm{D}$\xspace}
\newcommand{\FMu}{F$_\textrm{Mu}$\xspace}
\newcommand{\elec}{e$^-$\xspace}
\newcommand{\mup}{$\mu^+$\xspace} 
\newcommand{\Mum}{Mu$^-$\xspace}
\newcommand{\Mup}{Mu$^+$\xspace}
\newcommand{\Muz}{Mu$^0$\xspace}

\begin{document}

\preprint{APS/123-QED}

\title{Observation and manipulation of charge carrier distribution at the \SiOSi interface}

\author{Maria Mendes Martins}
\email{maria.martins@psi.ch}
\affiliation{%
 Laboratory for Muon Spin Spectroscopy, Paul Scherrer Institute, 5232 Villigen PSI, Switzerland
}
\affiliation{
  Advanced Power Semiconductor Laboratory, ETH Zurich,  8092 Zurich, Switzerland
}
 \author{Piyush Kumar}
 \affiliation{
  Advanced Power Semiconductor Laboratory, ETH Zurich,  8092 Zurich, Switzerland
}

\author{Marianne E. Bathen}
 \affiliation{
  Department of Physics/ Centre for Materials Science and Nanotechnology, University of Oslo, 0316 Oslo, Norway
}
\affiliation{
  Advanced Power Semiconductor Laboratory, ETH Zurich,  8092 Zurich, Switzerland
}

\author{Zaher Salman}
\affiliation{%
 Laboratory for Muon Spin Spectroscopy, Paul Scherrer Institute, 5232 Villigen PSI, Switzerland
}%

\author{Ulrike Grossner}
 \affiliation{
  Advanced Power Semiconductor Laboratory, ETH Zurich,  8092 Zurich, Switzerland
}
\author{Thomas Prokscha}%
 
\affiliation{%
 Laboratory for Muon Spin Spectroscopy, Paul Scherrer Institute, 5232 Villigen PSI, Switzerland
}%

\date{\today}

\begin{abstract}

Using low-energy muons, we map the charge carrier concentration as a function of depth and electric field across the \SiOSi interface up to a depth of \SI{100}{\nano\meter} in Si-based MOS capacitors. 
The results show that the formation of the anisotropic bond-centered muonium \MuBCz state in Si serves as a direct measure of the local changes in electronic structures.
Different band-bending conditions could be distinguished, and the extension of the depletion width was directly extracted using the localized stopping and probing depth of the muons. Furthermore, electron build-up on the Si side of the \SiOO/Si interface, caused by the mirror charge induced by the fixed positive charge in the oxide and the image force effect, was observed. 
Our work represents a significant extension of the application of the muon spin rotation technique ($\mu$SR) and lays the foundation for further research on direct observation of charge carrier density manipulation at technologically important semiconductor device interfaces.

\end{abstract}

\maketitle

Muon spin rotation (\muSR) spectroscopy provides valuable information about the electronic and magnetic environments of the material in which the positive muons (\mup) are implanted.
In the early days of \muSR, the \mup was used to understand the electronic structure and dynamics of hydrogen, an impurity commonly present in semiconductor materials \cite{Chow_1998musr, Lichti_2004, Cox_2009, muon_spectroscopy_2021}. More recently, additional external stimuli, such as photo-excitation \cite{Fan_2008,Prokscha_2013,Prokscha_2020, Lord_2020,Yokoyama_2017, Yokoyama_2021,Murphy_2022} and application of electric fields, allows shifting the focus to the investigation of carrier transport in semiconductors \cite{Storchak_1997,Storchak_2004,Eshchenko_2009}.
Typically, electrical characterization of semiconductor materials is performed with deep level transient spectroscopy (DLTS), capacitance-voltage (\textit{C-V}), and current-voltage (IV) measurements \cite{Schroder_2015_chap5_6,Johnson_1979}, to name a few. These techniques rely on the application of a varying voltage at the metal contact of a semiconductor device structure, causing band-bending at the metal-semiconductor interface. 
However, the existing energy-resolving electrical analysis tools have limited depth resolution, and are unsuitable to probe the narrow regions where electron and hole transport is directly influenced by the changes in band structure.

Low-energy muon spin spectroscopy (\LEmuSR) is a compelling extension of conventional \muSR, often used to study the narrow interface regions in thin-films and multilayer systems due to its nanometer depth  resolution \cite{Morenzoni_1994,Morenzoni_1997,Morenzoni_2000,Prokscha_2008}. By tuning the implantation energy of the muons between \SIlist{1;30}{\kilo\electronvolt}, it is possible to control their mean stopping depth in the range of \SIrange{5}{200}{\nano\meter}, such that even the near-surface and interface regions can be probed. This technique has been previously applied, e.g., to study charge carrier distributions at the surface of commercial germanium wafers \cite{Prokscha_2013,Prokscha_2020}, 
and to identify different species of point defects in semiconductors such as silicon carbide \cite{Woerle_2020,Martins_2023,Kumar_2024_AlImp}.
Here, we report the results obtained with a recently developed setup for performing \LEmuSR experiments to measure thin-film systems under the influence of an externally applied electric field (EF-LEM). For this, a metal-oxide-semiconductor (MOS) structure based on silicon (Si) was used, and the band-bending in Si is manipulated via the applied bias. The muonium states (Mu, bound state of a \mup and an \elec) in Si and the thermally activated transitions between them are well characterized \cite{Patterson_1988,Kreitzman_1995}. Their formation probabilities depend strongly on the type of majority charge carrier (hole or electron) and its concentration \cite{Hitti_1999}. In Si, muonium can exist at the center of a Si-Si bond --- the so-called bond-centered \MuBC~--- as \MuBCp or \MuBCz, or, if it is located at the tetrahedral site, as \MuTz or \MuTm \cite{Patterson_1988,Kreitzman_1995}. The diamagnetic (\Mup and \Mum) and paramagnetic (\Muz) states are spectroscopically distinguished by their different spin precession frequencies in an applied magnetic field, where
the muon spin will precess at the Larmor frequency in the diamagnetic states, while the spin precession frequencies in \Muz are different due to the hyperfine interaction between the \mup and \elec spins
\cite{muon_spectroscopy_2021,hillier_muon_2022}. The formation of the final Mu charge state is also affected by the implantation process; the \MuBCz is likely to form by \textit{delayed} capture of a track \elec (from the ionization track generated by the slowing down process of \mup) \cite{Krasnoperov_1992,Eshchenko_2002,Eshchenko_2002a,Prokscha_2007}. In the case of LE-\mup, the total number of track \elec depends on the implantation energy of the incoming \mup beam, and energies $>$\SI{10}{\kilo\electronvolt} are required to efficiently form \MuBCz \cite{Prokscha_2007}.
Herein, due to its sensitivity to the presence of electrons \cite{Patterson_1988}, the anisotropic \MuBC state in Si was used as a charge carrier probe at the interface of a MOS structure. 
With EF-LEM, we could, for the first time, directly probe and manipulate the depth-dependence of the charge carrier concentration due to band bending modification at the Si surface. The main findings are observed when an upward bending in n-type Si bands is induced, which enables probing of depletion and inversion layers. Furthermore, an estimate of the space charge width evolution with reverse bias could be obtained.

The experiments were conducted in the low-energy muon (LEM) instrument of the Swiss Muon Source (S$\mu$S) at the Paul Scherrer Institute, Villigen, Switzerland. The MOS sample measured consisted of \SI{20}{\nano\meter} aluminum (Al) and \SI{50}{\nano\meter} of silicon dioxide (\SiOO) deposited on a commercial wafer of n-type Si. The \SiOO layer was deposited on Si(100) with plasma-enhanced chemical vapor deposition (PECVD) at \SI{300}{\celsius}.  The Si wafer was doped with phosphorus (P) to a concentration N$_\textsc{D}=$~\SI{3e18}{\per\cubic\centi\meter}.

\begin{figure}[hbt!]
\includegraphics[width=0.49\textwidth]{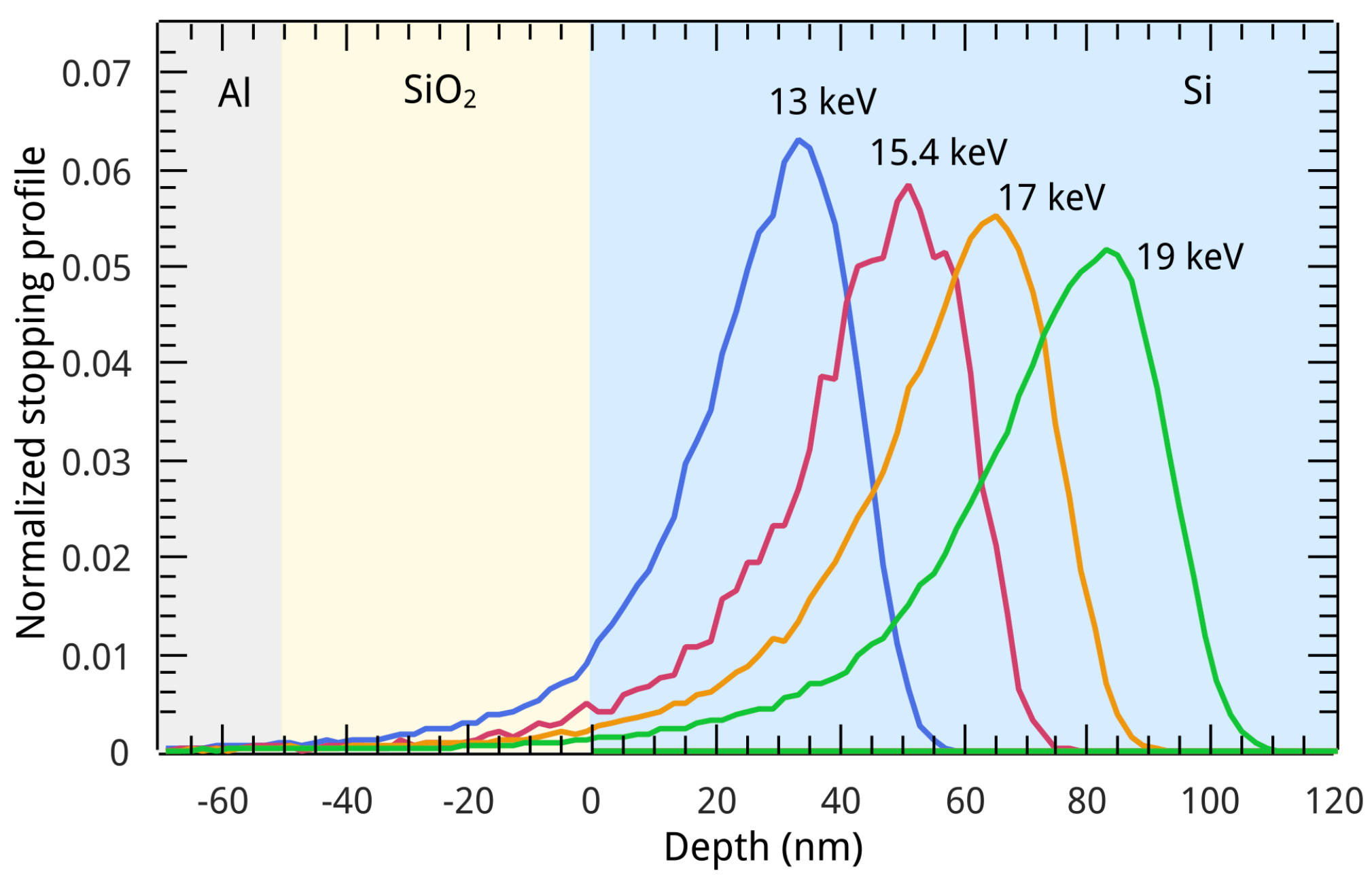}
\vspace{-0.3cm}
\caption{\label{fig:stopProfiles}Stopping distribution of the muons in the Al/\SiOO/Si MOS structure. The profiles of the measured implantation energies were obtained from Monte Carlo simulations in TRIMSP \cite{Eckstein_1991,Morenzoni_2002}.}
\end{figure}

Spin polarized muons were implanted into the Si MOS structure. For all the measurements, a transverse magnetic field (TF, perpendicular to the \mup spin) was applied. To probe the semiconductor near the \SiOSi interface, the muon implantation energy was varied in the range \SIrange{13}{19}{\kilo\electronvolt}. The stopping profiles of the muons for each implantation energy were simulated using the program TRIMSP \cite{Eckstein_1991,Morenzoni_2002}, and are shown in \Cref{fig:stopProfiles}.

\begin{figure}[hbt!]
\includegraphics[width=0.45\textwidth]{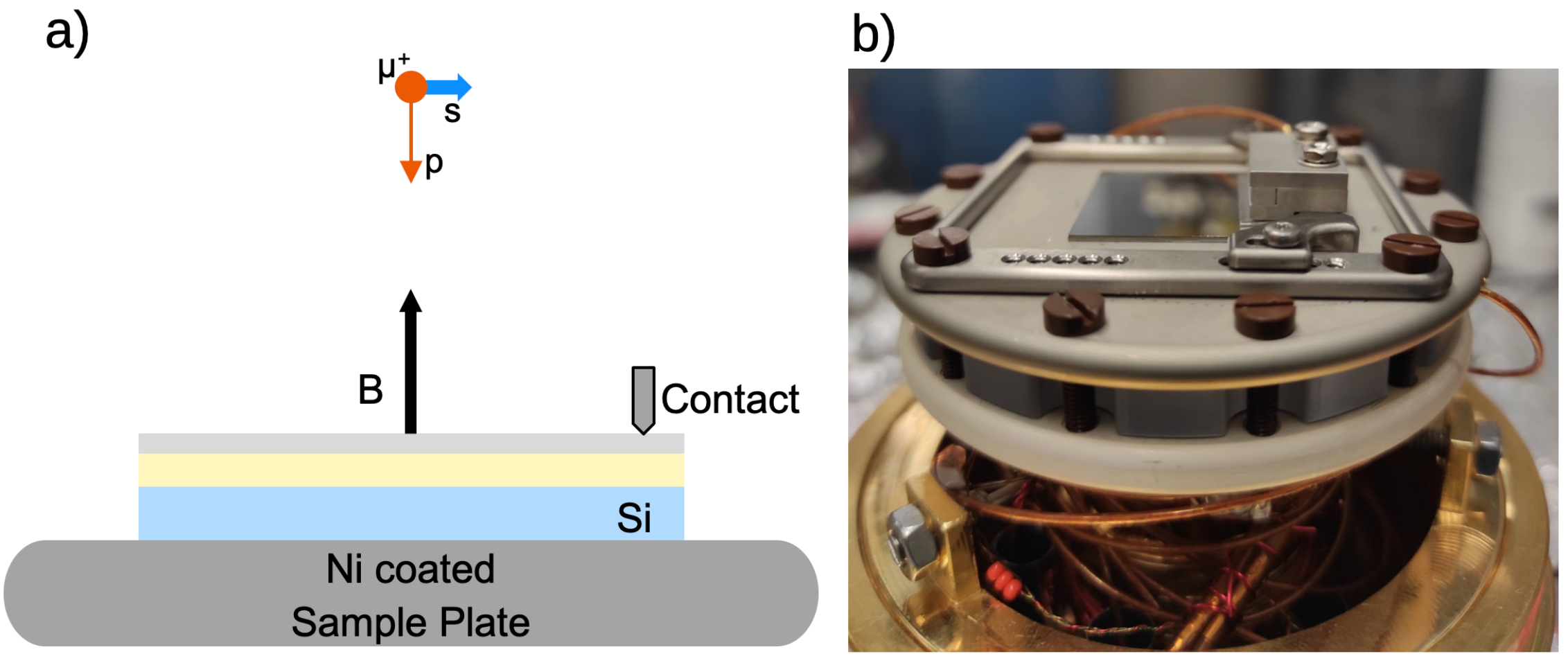}
\caption{\label{fig:EFsetup}a) Schematic representation of the experimental setup of the EF-LEM experiment. \textbf{s} denotes the spin, and \textbf{p} the momentum of the \mup. b) Sample plate with Si MOS samples mounted on the cryostat. A frame is placed on the sample plate and fixed with screws. On top of the sample, spring-loaded connectors are used to contact the Al layer of the MOS.}
\end{figure}

The electric field (EF) setup consisted of a frame with spring-loaded connectors that is placed on top of the sample in contact with the metal layer of the MOS capacitor. An electric field is applied perpendicular to the sample surface (see \Cref{fig:EFsetup}). All beam-exposed parts of the frame and the sample plate are coated with a \SI{500}{\nano\meter} layer of ferromagnetic nickel (Ni), to suppress muon-spin precession contributions from muons stopping outside the sample (the recorded background decay asymmetry of muons stopping in the Ni layer is A(Ni)$\leq0.01$  in the relevant energy range \cite{Suter_low_2023}). The \mup implantation energy is primarily changed by altering the high voltage applied to the sample plate to accelerate or decelerate the incoming muons, whilst a low voltage between \SIrange{-5}{2}{\volt} is applied to the top electrode of the MOS capacitor to induce an electric field across the oxide layer. Thus, to perform EF measurements while probing different depths in the material, the sample and the connecting devices, such as the power supply, are floated.

The EF is applied parallel to the direction of the incoming muon momentum. This means that the free electrons in Si are attracted towards the \SiOSi interface when a positive bias is applied to the Al top contact (gate terminal) of the MOS capacitor. On the other hand, a reverse gate bias causes the free electrons to be pushed away from the interface and deeper into the Si. Since \MuBCz quickly depolarizes by spin-exchange reactions with free electrons at concentrations $>$\SI{1e13}{\per\cubic\centi\meter} \cite{Patterson_1988}, the presence of free electrons causes a significant broadening of the \MuBCz muon spin precession lines, resulting in the disappearance of the \MuBCz precession signal in the $\mu$SR spectra.
Thus, EF can change the observable \MuBCz fraction in the $\mu$SR spectra and \MuBCz can be used as a very sensitive probe to the presence of free electrons in the surroundings of the implanted muon.

\begin{figure}[hbt!]
\includegraphics[width=0.5\textwidth]{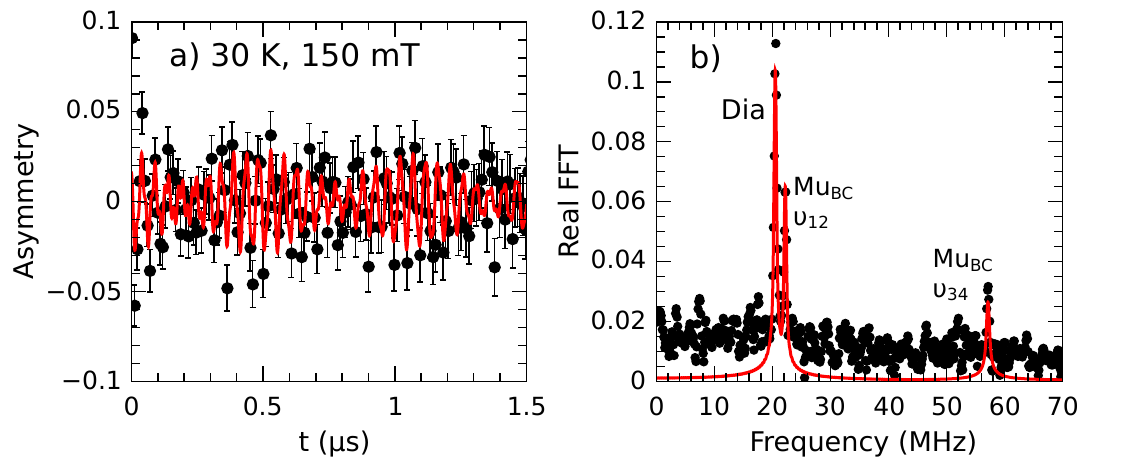}

\caption{\label{fig:Asym_FFT}a) \muSR asymmetry time spectrum measured 
at \SI{30}{\kelvin} and \SI{150}{\milli\tesla} and b) corresponding fast Fourier transform (Real FFT). The peak at \SI{20.49(1)}{\mega\hertz} corresponds to the diamagnetic precession frequency. $\nu_{12}$ and $\nu_{34}$ are the high-field transition lines of \MuBCz, with frequencies of \SIlist{22.20(1);57.14(7)}{\mega\hertz}, respectively.}
\end{figure}

The temperature dependence of \MuBCz formation in Si was measured between 30~K and 90~K, before any electric field was applied. For these initial measurements, the implantation energy was \SI{15.4}{\kilo\electronvolt} with zero bias at the sample plate, which corresponds to a mean \mup implantation depth $\langle x \rangle$ of \SI{42}{\nano\meter} in Si beneath the \SiOSi interface. All the results were obtained at a magnetic field $\textbf{B}=$ \SI{150}{\milli\tesla}. The \muSR asymmetry time spectra were fitted using the program {\tt musrfit} \cite{Suter_2012}
(as shown in \Cref{fig:Asym_FFT}~a)) to determine the asymmetries of the diamagnetic (A$_\mathrm{Dia}$) and paramagnetic (A$_{\mathrm{Mu}}$) signals, where A$_{\mathrm{Mu}}$ denotes the sum of asymmetries of the \MuBCz precession lines. 
%
%
The paramagnetic \FMu and diamagnetic \FD fractions were calculated as F$_\textrm{Mu/D}$~$=\text{A}_\textrm{Mu/D}\div \text{A}_{\text{Ag}}$.  
\text{A}$_{\text{Ag}}$ is the maximum asymmetry of the LEM setup measured in silver, where there is almost no depolarization of the muon signal \cite{SAADAOUI_2012}.
At \SI{150}{\milli\tesla}, the precession frequencies of \MuTz are too high to be resolved with the limited time resolution of the LEM instrument, while two ``high-field'' transition lines of \MuBCz,
$\nu_{12}$ and $\nu_{34}$ with frequencies \SI{22.20(1)}{\mega\hertz} and \SI{57.14(7)}{\mega\hertz}, respectively, are observable.
As shown in \Cref{fig:Asym_FFT}~b), when \MuBCz forms, three lines are present in the frequency domain; the diamagnetic and the two \MuBCz lines. This is the case at low temperatures up to \SI{30}{\kelvin}.
Once the ionization of the phosphorus donors starts at T~$>$~\SI{30}{\kelvin}, \MuBCz depolarizes due to fast spin exchange between the bound electron and the free electrons \cite{Patterson_1988,Senba_1994}, and the paramagnetic signal is no longer visible. 
%
%

\begin{figure}[hbt!]
\includegraphics[width=0.48\textwidth]{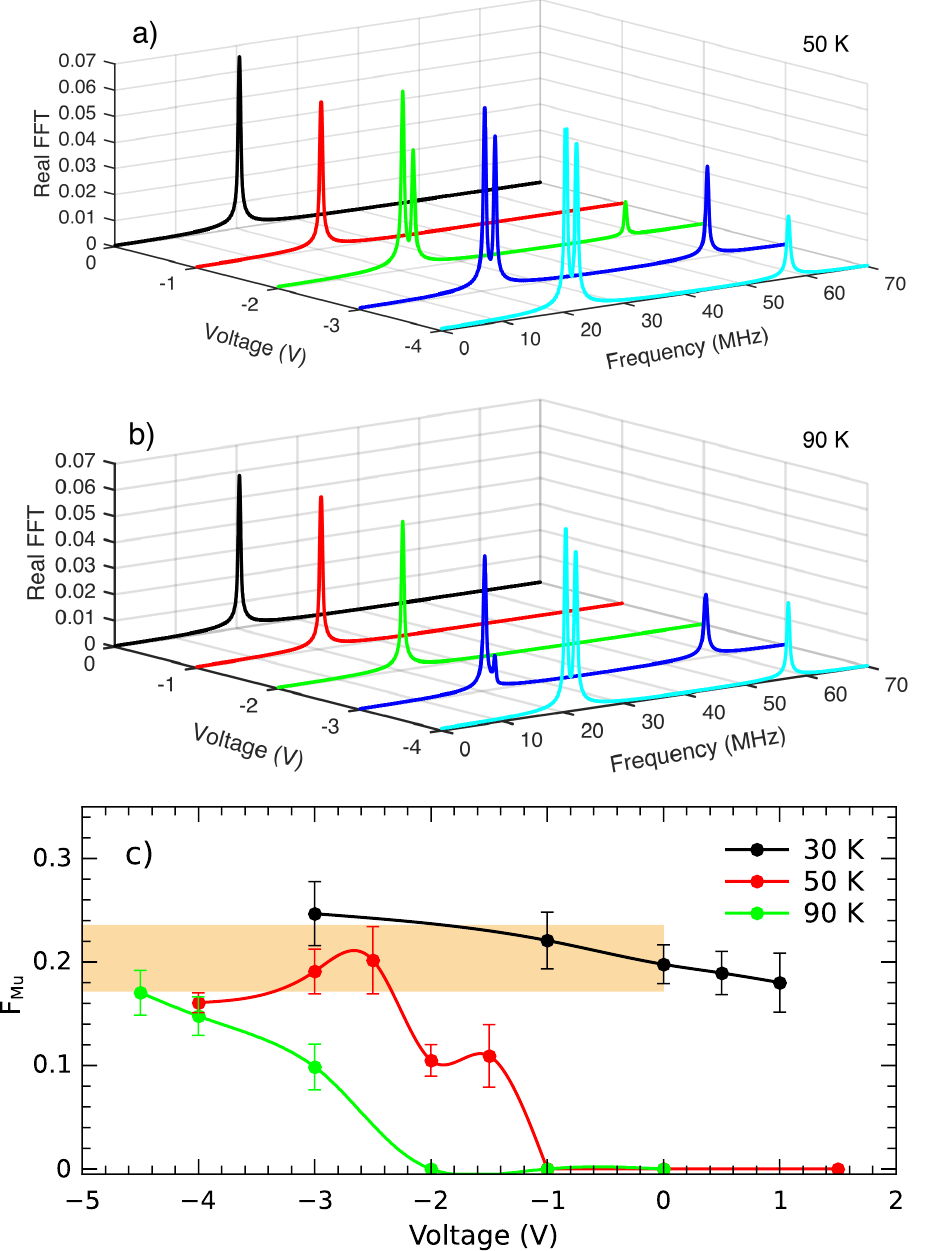}

\caption{\label{fig:FFTBiasScan}Fast Fourier transforms (real FFT) of the asymmetry spectra for different applied biases at an implantation energy of \SI{15.4}{\kilo\electronvolt} at a) \SI{50}{\kelvin} and b) \SI{90}{\kelvin}. c) Fraction \FMu of implanted muons forming the paramagnetic \MuBCz state as a function of the applied negative bias. The lines are guides to
the eyes. The orange region corresponds to the \FMu level in depletion condition. }
\end{figure}

Electric field measurements at \SI{50}{\kelvin} and \SI{90}{\kelvin}, in the presence of free electrons ($n>$~\SI{3e16}{\per\cubic\centi\meter}), show the reappearance of the \MuBCz precession lines (see \Cref{fig:FFTBiasScan}~a)-b)) when a reverse gate bias is applied. This points to a reduction of electron concentration to $n <$~\SI{1e13}{\per\cubic\centi\meter} \cite{Patterson_1988} in the probed region ($\langle x \rangle=$ \SI{42}{\nano\meter} in Si) as the applied gate voltage induces electron depletion.
At \SI{50}{\kelvin}, a gate voltage of \SI{-2}{\volt} is enough to induce electron depletion, while at \SI{90}{\kelvin} the \MuBCz lines only reappear at \SI{-3}{\volt} gate bias in the same region. This reflects the temperature-dependent behavior of the donor activation, and increasing concentration of free electrons between \SI{50}{\kelvin} (n~$\approx$~\SI{5e16}{\per\cubic\centi\meter}) and \SI{90}{\kelvin} (n~$\approx$~\SI{2e17}{\per\cubic\centi\meter})\cite{Grundmann_2010}.
The \FMu measured as function of applied bias is summarized in \Cref{fig:FFTBiasScan}~c) in the presence  (\SI{50}{\kelvin} and \SI{90}{\kelvin}) and absence (\SI{30}{\kelvin}) of free electrons. When a positive bias was applied, no effect on the \muSR signal was observed. This is expected due to the freeze-out of carriers at \SI{30}{\kelvin}, and the formation of an electron accumulation layer, which contributes to the depolarization of the \MuBCz state at \SI{50}{\kelvin} and \SI{90}{\kelvin}.

%
%

\begin{figure}[hbt!]
\includegraphics[width=0.48\textwidth]{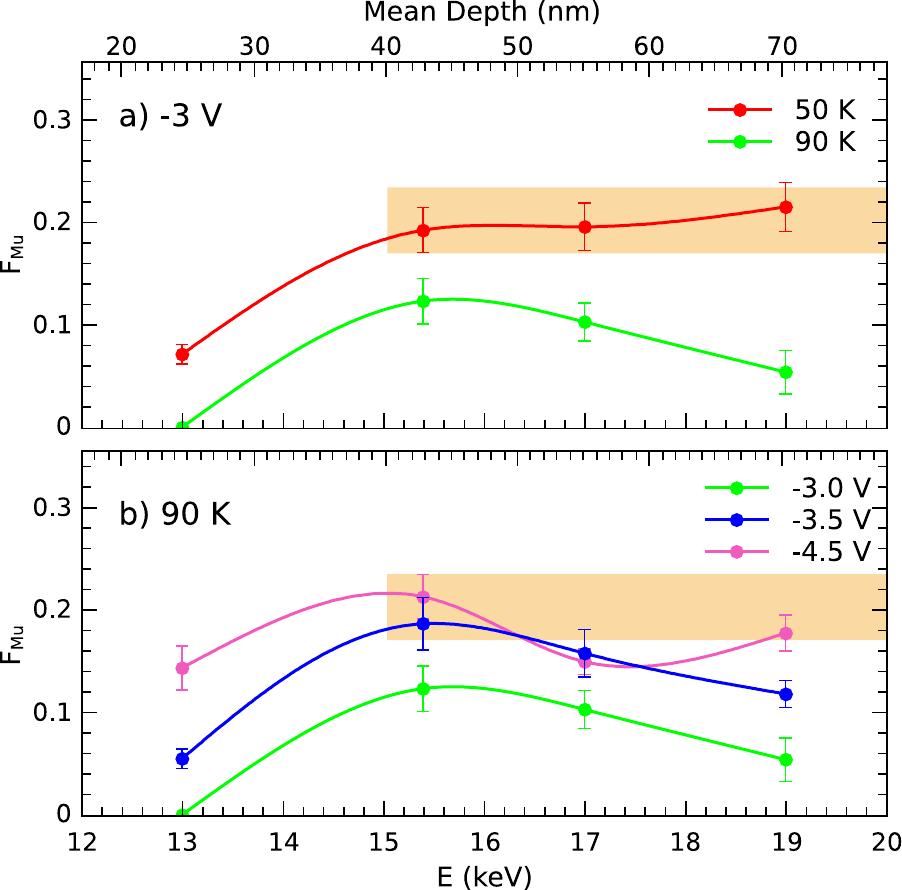}
 \caption{\label{fig:MuFracEscan}\FMu measured as a function of muon implantation energy, a) for an applied gate bias of \SI{-3}{\volt} at \SIlist{50;90}{\kelvin}, and b) at \SI{90}{\kelvin} when a reverse gate bias in the range \SIrange{-3}{-4.5}{\volt} is applied. The lines are guides to the eyes. }
\end{figure}

The difference in free carrier availability at \SI{50}{\kelvin} and \SI{90}{\kelvin} is also observable as a function of depth in \Cref{fig:MuFracEscan}~a). At \SI{-3}{\volt}, the \MuBCz lines are present for both temperatures at $\langle x \rangle >$~\SI{30}{\nano\meter} due to electron depletion. However, the applied voltage leads to the depletion region being wider than \SI{100}{\nano\meter} at \SI{50}{\kelvin}, as shown by \FMu values in the orange area, and around \SI{50}{\nano\meter} at \SI{90}{\kelvin}. 

\begin{figure}[hbt!]
\includegraphics[width=0.48\textwidth]{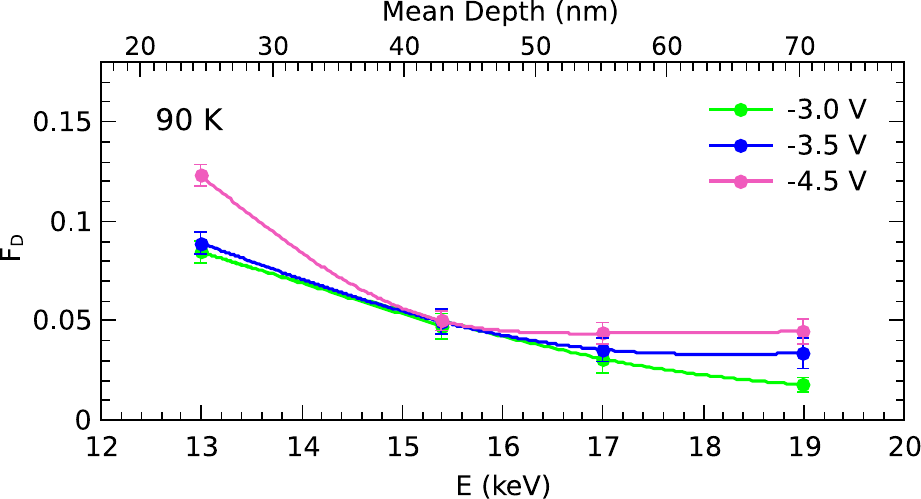}
 \caption{\label{fig:DiaFracEscan}\FD measured as a function of muon implantation energy, at \SI{90}{\kelvin} when a reverse gate bias is applied. The lines are guides to the eyes. }
\end{figure}

At \SI{90}{\kelvin}, the effect of the applied bias at different depths in Si is visible in \Cref{fig:MuFracEscan}~b) and \Cref{fig:DiaFracEscan}. This provides information about the evolution of depletion region width and the progressive change of the charge density near the \SiOSi interface.
The applied gate bias (V~$<$~\SI{0}{\volt}) starts to induce the formation of a depletion region between \SIlist[list-units=single]{35;70}{\nano\meter}, indicated by the gradual increase of \FMu. The depletion layer extends to a mean depth of \SI{55}{\nano\meter} at \SI{-3}{\volt}, \SI{65}{\nano\meter} at \SI{-3.5}{\volt}, and \SI{70}{\nano\meter} at \SI{-4.5}{\volt}.
It has been reported that the \SiOO deposited with the PECVD process has high positive fixed charge density \cite{Yasutake_1998}, creating a thin electron accumulation layer in the presence of free electrons, which is considered to be significant for the depletion to weak inversion regime \cite{bonilla_charge_2020}. At \SI{13}{\kilo\electronvolt}, between the interface and a mean depth of \SI{25}{\nano\meter}, an electron build-up induced by the positive oxide fixed charge \cite{Copuroglu_band-bending_2013} is sensed by the reduced \MuBCz formation, compared to the point at \SI{15}{\kilo\electronvolt}.
 It is also possible that the image-force effect plays a role in altering the charge distribution near the \SiOSi interface. In a Schottky diode, image-force is known to induce the lowering of the barrier for carrier emission \cite{Sze_2007}. In the context of the MOS capacitor, when a charge in the semiconductor is near the oxide interface, it induces an image charge on the metal gate. This results in an attractive force pulling the charge carrier towards the oxide-semiconductor interface \cite{Gehring_2006tunneling,Wen_1998}, and could explain the increase in electron concentration in the region probed at \SI{13}{\kilo\electronvolt}.
The electron concentration decreases with increasing reverse bias, and the \FMu value associated with the depletion regime is reached at \SI{-4.5}{\volt}.
At \SI{-4.5}{\volt} the increase in \FD is explained by the formation of an inversion layer (see \Cref{fig:DiaFracEscan}). It is proposed that the increase in hole concentration leads to hole capture by \MuTz and the formation probability of \Mup is enhanced. It is further proposed that the inversion layer does not lead to depolarization of the \MuBCz because a concentration of holes 100$\times$ higher than electrons would be required, which would be on the order of \SI{1e15}{\per\cubic\centi\meter} \cite{Patterson_1988}.

Capacitance-voltage (\textit{C-V}) measurements at a frequency of \SI{10}{\kilo\hertz} were performed to compare with the \LEmuSR results. During a \textit{C-V} measurement, a bias is applied to the metal gate and swept in steps, to vary the space charge region and band-bending, and the capacitance as a function of the applied voltage is recorded. 

\begin{figure}[tb!]

\includegraphics[width=0.5\textwidth]{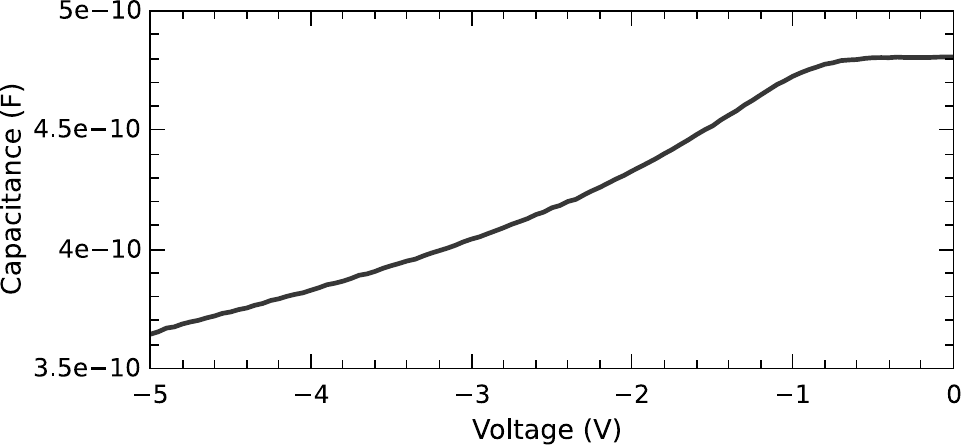}
\caption{\label{fig:CVcurve}\textit{C-V} curve measured at \SI{300}{\kelvin} at a frequency of \SI{10}{\kilo\hertz}. }
\end{figure}

 The \textit{C-V} curve in \Cref{fig:CVcurve} was obtained at \SI{300}{\kelvin}, when the ionization of donors is expected to be complete, and the free carrier concentration is equal to the doping concentration. The MOS depletion width ($W_D$) values at each temperature (\SIlist[list-units=single]{50;90}{\kelvin}) for the relevant voltage points, calculated based on the measured \textit{C-V}, are shown in \Cref{tab:width}.
At \SI{50}{\kelvin} and \SI{-3}{\volt}, the electron depletion extends beyond the \LEmuSR probed region, meaning $W_D >$\SI{100}{\nano\meter}, in line with the calculated value. 
 A left shift of the \textit{C-V} curve is observed, indicating the presence of positive fixed oxide charges \cite{Sze_2007}, and in agreement with the interpretation of the \LEmuSR data that a region with a negative charge is induced in Si near the interface. 
 At \SI{-1.1}{\volt} the onset of depletion can be viewed in \Cref{fig:CVcurve}, while complete electron depletion at \SI{50}{\kelvin} is only observed with \LEmuSR at \SI{-3}{\volt}. The electron concentration likely starts to decrease in the region between \SIrange{30}{50}{\nano\meter} at \SI{-1}{\volt}, but only at \SI{-3}{\volt} it gets lower than \SI{1e13}{\per\cubic\centi\meter} in the entire region. At V~$<$~\SI{-3}{\volt} the calculated $W_D$ matches well the \LEmuSR observations for the width of the electron depletion region at \SIlist[list-units=single]{50;90}{\kelvin}.

\begin{table}[bt!]
    \centering
    \caption{Depletion width $W_D$ calculated from the \textit{C-V} curve at \SI{300}{\kelvin} and the EF-LEM measurements at \SIlist{50;90}{\kelvin}.}
    \begin{tabular}{c|cccc} 
        \multirow{ 3}{*}{Voltage (V)}  &  \multicolumn{4}{c}{$W_D$ (nm) \hspace{5mm}} \\ 
          &  \multicolumn{2}{c}{\textit{C-V} \hspace{5mm}}&  \multicolumn{2}{c}{EF-LEM \hspace{5mm}} \\ 
          &\hspace{2.2mm}50~K \hspace{2.5mm} &\hspace{2.2mm}90~K &\hspace{2.2mm}50~K\hspace{2.5mm} &\hspace{2.2mm}90~K \hspace{2.5mm}\\ \hline
         -3.0 V  &\SI{145}{}&\SI{56}{}& $>$\SI{100}{}  & \SI{55}{}   \\ 
         -3.5 V  & \SI{197}{}& \SI{61}{}& ---  & \SI{65}{}\\ 
         -4.5 V & \SI{228}{}  & \SI{71}{}& ---  & \SI{70}{} \\ \hline
    \end{tabular}
    \label{tab:width}
\end{table}

In this work, charge carrier concentration changes induced by accumulation, depletion, and inversion conditions in Si, in the vicinity of the \SiOO/Si interface, were probed with electrical field LE-$\mu$SR.
Similar to \textit{C-V} measurements, different band-bending conditions could be distinguished. However, additional information about the extension of the depletion width, from the interface to a mean depth of \SI{70}{\nano\meter}, as a function of the applied gate bias, was directly extracted due to the localized stopping and probing depth of the muon.
Interestingly, we can observe the mirror charge induced by the fixed positive charge in the oxide and image force effect, in the form of electron build-up on the Si side of the \SiOO/Si interface.

Overall, we demonstrate that this technique can be used to study functional semiconductor device structures, acquiring novel local/nanometer-scale information and a better understanding of the physical mechanisms that affect the charge carrier dynamics at critical regions (e.g. oxide-semiconductor interface). As processing steps such as post-oxidation annealing are known to affect the electronic character of the oxide/semiconductor interface (see, for example, \cite{Kumar_2023}), this further supports the importance and applicability of muon-probed, nanometer-resolved manipulation of charge carrier concentration.

The muon measurements have been performed at the Swiss Muon Source S$\mu$S, Paul Scherrer Institute, Villigen, Switzerland. This work is supported by the Swiss National Science Foundation (SNSF) by grant number 200021192218. 
The work of MEB was supported by an ETH Z{\"u}rich Postdoctoral Fellowship.

The raw data that support the findings of this study are openly available at the following URL/DOI: \url{doi.psi.ch/detail/10.16907/9543c23c-7282-44d0-8b47-74ea18557954}. 

\bibliographystyle{apsrev4-1}
\bibliography{library}

\end{document}